\documentclass[aps,prd,amsmath,nofootinbib,twocolumn,10pt,showpacs,preprintnumbers]{revtex4-1}
\usepackage{graphicx}
\usepackage{color}
\usepackage[retainorgcmds]{IEEEtrantools}
\usepackage{hyperref}
\usepackage{dcolumn}
\usepackage{multirow}

\newcommand{\code}[1]{\texttt{#1}}
\newcommand{\be}{\begin{equation}}
\newcommand{\ee}{\end{equation}}
\newcommand{\bea}{\begin{eqnarray}}
\newcommand{\eea}{\end{eqnarray}}
\newcommand{\bit}{\begin{itemize}}
\newcommand{\eit}{\end{itemize}}
\newcommand{\bfi}{\begin{figure}}
\newcommand{\efi}{\end{figure}}
\newcommand{\bfic}{\begin{figure*}}
\newcommand{\efic}{\end{figure*}}
\newcommand{\bce}{\begin{center}}
\newcommand{\ece}{\end{center}}
\newcommand{\bt}{\begin{table}}
\newcommand{\et}{\end{table}}
\newcommand{\btb}{\begin{tabular}}
\newcommand{\etb}{\end{tabular}}

\newcommand{\highpsis}{$\Psi_2$, $\Psi_3$ and $\Psi_4$}
\newcommand{\lowpsis}{$\Psi_0$ and $\Psi_1$}

\begin{document}

\title{Falloff of the Weyl scalars in binary black hole spacetimes}

\author{Ian Hinder}
\email{ian.hinder@aei.mpg.de}
\affiliation{
Max-Planck-Institut f\"ur Gravitationsphysik,
Albert-Einstein-Institut, \\ 
Am M\"uhlenberg 1, D-14476 Golm, Germany}

\author{Barry Wardell}
\email{barry.wardell@aei.mpg.de}
\affiliation{
Max-Planck-Institut f\"ur Gravitationsphysik,
Albert-Einstein-Institut, \\ 
Am M\"uhlenberg 1, D-14476 Golm, Germany}

\author{Eloisa Bentivegna}
\email{eloisa.bentivegna@aei.mpg.de}
\affiliation{
Max-Planck-Institut f\"ur Gravitationsphysik,
Albert-Einstein-Institut, \\ 
Am M\"uhlenberg 1, D-14476 Golm, Germany}

\preprint{AEI-2011-025}

\date{\today}

\begin{abstract}
  The peeling theorem of general relativity predicts that the Weyl curvature scalars $\Psi_n$ ($n=0,\dots,4$), when constructed from a suitable null tetrad in an asymptotically flat spacetime,
  fall off asymptotically
  as $r^{n-5}$ along outgoing radial null geodesics.  This leads to the interpretation of $\Psi_4$ 
  as outgoing gravitational radiation at large distances from the source. 
  We have performed numerical simulations in full
  general relativity of a binary black hole inspiral and merger, and have computed
  the Weyl scalars in the standard tetrad used in numerical relativity. In
  contrast with previous results~\cite{Pollney:2009ut}, we observe that all
  the Weyl scalars fall off according to the predictions of the theorem.
\end{abstract}
\pacs{04.25.dg, 04.20.Ha, 04.30.Db}

\maketitle

\section{Introduction}

Spacetimes containing multiple black holes (BHs)
are perhaps the most studied class of non-perturbative, low-symmetry vacuum solutions of
Einstein's equations and have already opened the path to tests of the
generality of a number of fundamental theorems and conjectures in general relativity.

Studies of multiple-BH systems have addressed the behaviour of trapped surfaces and 
tubes~\cite{Schnetter:2004mc,Campanelli:2006fg,Campanelli:2006fy,
Schnetter:2006yt,Szilagyi:2006qy,Jasiulek:2009zf,Owen:2009sb} and event 
horizons~\cite{Matzner:1995ib, Masso:1998fi,Caveny:2003er,Diener:2003jc,
Alcubierre:2004hr,Cohen:2008wa,Ponce:2010fq} during the dynamical 
many-body phase.
The asymptotic properties (such as the radial falloff of the Weyl
scalars) of binary-black-hole (BBH) spacetimes have been investigated~\cite{Pollney:2009ut},
and the algebraic character of BBH spacetimes has been studied~\cite{Campanelli:2008dv,Owen:2010vw}. 
A program to include conformal infinity in the numerical domain
has been carried out~\cite{Reisswig:2009rx,Reisswig:2009us,Babiuc:2010ze},
and properties of higher-dimensional BBH systems have been described~\cite{Zilhao:2010sr,Witek:2010xi,Witek:2010az}.
Several interesting directions remain to be explored in the context of multiple-BH
evolutions~\cite{Karkowski:2004gh,Jaramillo:2007mi,Dain:2009qb,Mars:2009cj,
Lehner:2007ip,Gallo:2008pm}.

It seems fair to state that three-dimensional numerical evolutions of 
Einstein's equations are now leading to compelling and useful results for
gravitational fields in 
non-trivial configurations.  Whilst this class of 
numerical relativity (NR) studies may appear somewhat removed from
the field of gravitational-wave astrophysics, the clarification of fundamental
issues has a direct impact there. For instance,
the interpretation of $\Psi_4$ as the outgoing component of 
gravitational radiation relies entirely on the applicability of the 
peeling theorem.

In this work, we follow up the study carried out in~\cite{Pollney:2009ut},
where the radial falloff of the Weyl scalars was
measured in a BBH spacetime and compared with the results
expected from the peeling theorem.  We use the same evolution code and the same definitions of
the Weyl scalars.  However, we use an independent
and open analysis framework provided by the Einstein Toolkit
initiative~\cite{EinsteinToolkit:web} for computing the scalars\footnote{The unexpected
results pointed out in~\cite{Pollney:2009ut} for the peeling properties of $\Psi_0$ and
$\Psi_1$ were affected by errors in the original analysis implementation which
have subsequently been corrected.
As an aside, we note that ensuring correctness is a subtle and
involved part of developing complex scientific applications~\cite{Sargent:1999:VVS:324138.324148}. We
believe that collaborative development and the use of open-source
software is one of the simplest and most effective ways to ensure
code correctness.
}, and we analyse their falloff properties using a more
direct approach. Our results indicate that all of the
$\Psi_n$ fall off at the expected rates, suggesting that the usual techniques used
for wave extraction in NR simulations are sufficient for
recovering the results of the peeling theorem. We present our method in 
Sec.~\ref{sec:methodology} and the results we obtain in Sec.~\ref{sec:results}.
We conclude with a discussion in Sec.~\ref{sec:discussion}.
Throughout this paper, we use a space-like signature $(-,+,+,+)$ and
a system of units in which $c=G=1$.

\section{Methodology}
\label{sec:methodology}

\subsection{The Weyl scalars and the peeling theorem}
\label{sec:weyl}

The work of Sachs~\cite{Sachs:1961zz}, Newman and Penrose~\cite{Newman:1961qr} illustrates how the information encoded in 
the traceless part of the Riemann curvature tensor, the Weyl tensor $C_{abcd}$, can be completely expressed 
as a set of five complex numbers usually referred to as the \emph{Weyl scalars}:
\begin{subequations}
\bea
\Psi_0 &=& C_{abcd} \ell^a m^b \ell^c m^d \\
\Psi_1 &=& C_{abcd} \ell^a n^b \ell^c m^d \\
\Psi_2 &=& C_{abcd} \ell^a m^b \bar m^c n^d \\
\Psi_3 &=& C_{abcd} \ell^a n^b \bar m^c n^d \\
\Psi_4 &=& C_{abcd} n^a \bar m^b n^c \bar m^d
\eea
\end{subequations}
where $(n^a, \ell^a, m^a, \bar{m}^a)$ is a tetrad of two real and
two complex null vectors satisfying $n_a n^a = 0$, $\ell_a \ell^a=0$,
$n_a \ell^a=-1$, $m_a m^a=0$ and $m_a \bar m^a = 1$.

Handling the curvature in this notation has a number of benefits: 
from the geometric standpoint, there is immediate insight to be gained 
as the five $\Psi_n$ are simply the spin-frame components of the Weyl spinor
$\Psi_{ABCD}$. Thus if the spin-frame has some physical relevance
(e.g.~is oriented along principal null directions), then the $\Psi_n$
can be identified as specific (radiative vs.~Coulomb, transverse vs.~longitudinal) 
components of the gravitational field~\cite{Szekeres:1965ux}.
From the practical side, the five complex quantities above simplify the
analysis of the asymptotic properties of the Riemann tensor (see, e.g.,%
~\cite{Newman:1961qr,Newman:1962}). In particular, the work of Newman 
and Penrose shows how these scalars fall off
along outgoing radial null geodesics in a neighbourhood of future
null infinity, $\cal J^+$, as:
\be
\Psi_n \sim r^{n-5}.
\ee 
(For the parallel result on $\cal J^-$, see e.g.~\cite{Walker:1979zk}.)
The usual assumptions of the Penrose conformal construction are
assumed, namely the existence of a spacetime conformally related to
the physical one by $g_{ab}^{\rm unphys}=\Omega^2 g_{ab}^{\rm phys}$,
with $\Omega=0$ at $\cal J^+$, the falloff of the corresponding 
Weyl tensor at least as $C_{abcd}^{\rm unphys} \sim \Omega$, also at $\cal J^+$,
and the alignment of $n^a$ and $\ell^a$ with the ingoing and outgoing
null directions, respectively (accomplished, for instance, by setting
$n_a = \Omega_{;a}$).
In the above, $r \equiv \Omega^{-1}$ can be used, at least in a neighbourhood 
of $\cal J^+$, as the affine parameter on the null geodesic.  In typical
asymptotically-flat spacetimes, this can be identified with a
radial coordinate, as long as the Weyl tensor satisfies the above 
falloff condition when expressed in terms of it.
This result, known as the peeling theorem, is a convenient tool for 
evaluating integrals at $\cal J^+$ and allows one to identify 
$\Psi_4$ with the outgoing gravitational-wave degrees of freedom since,
for one, it is the only component with an associated flux at
$\cal J^+$ that is not identically zero:
\be
\int_{\cal J^+} |\Psi_n|^2 dS \ne 0 \Rightarrow n=4.
\ee
Notice that, whilst this result is rather robust and valid in a
large class of tetrad (or spinor) frames, it does rely on a suitable
choice of $r$.
A question that was recently raised in~\cite{Pollney:2009ut} is to what extent 
the peeling theorem can be directly applied to the Weyl scalars usually calculated 
in NR simulations of BBH systems, 
where a dynamically-evolving gauge can obscure the character of the coordinates,
from which the tetrad is usually constructed.
This question is closely related to similar 
investigations carried out in the NR literature, such as what 
conditions should be imposed on the tetrad frame in order to retain the 
interpretation of $\Psi_4$ as outgoing gravitational waves~\cite{Lehner:2007ip}, or how many
principal null directions exist in post-merger black hole remnants~\cite{Campanelli:2008dv,Owen:2010vw}.

In this work, we investigate the extent to which the falloff rates predicted 
by the peeling theorem apply to BBH spacetimes as usually computed in
NR.  There are two aspects to this question: whether a numerical spacetime
satisfies the theorem's physical requirements
(the falloff condition for the Weyl tensor), and whether the dynamical
coordinates are compatible with the assumptions.

\subsection{Tetrad}
\label{sec:tetrad}

In numerical simulations, where the domain does not typically include
$\cal J^+$ (the notable exception being simulations using characteristic
extraction~\cite{Reisswig:2009rx,Reisswig:2009us}),
the tetrad $(n^a, \ell^a, m^a, \bar{m}^a)$ is
usually chosen according to the following straightforward prescription~\cite{Baker:2001sf}. First,
one defines the spatial vectors $\phi^a$, $r^a$ and $\theta^a$ via their spatial
components:
\begin{subequations}
\begin{IEEEeqnarray}{rCl}
{\phi}^i &=& [-y, x, 0], \\
{r}^i &=& [x, y, z], \\
{\theta}^i &=& \sqrt{\gamma} \epsilon^i{}_{jk} \phi^j r^k.
\end{IEEEeqnarray}
\end{subequations}
where $\gamma$ is the determinant of the spatial metric.
One then uses Gram-Schmidt orthonormalisation to produce the orthonormal triad
\begin{gather}
 e_\phi^a =\frac{{\phi^a}}{\| {\phi} \|},  \quad
 e_r^a = \frac{{r^a}-P_{ e_\phi}{r^a}}{\|{r}-P_{ e_\phi}{r}\|}, \nonumber\\
 e_\theta^a = \frac{{\theta^a} - P_{ e_\phi}{r^a} - P_{ e_r}{\theta^a}}{\|{\theta} - P_{ e_\phi}{r} - P_{ e_r}{\theta}\|},
\end{gather}
where $P_{ x}  y^a \equiv (  x_b y^b ) x^a$ is the projection of a vector $y^a$ along a
unit vector $x^a$, $\| v\|=\sqrt{ v_a v^a}$, and indices are raised and lowered
with the spatial metric. Finally, 
one complements this triad with the unit hypersurface normal $u^a$ and
constructs the null tetrad as
\begin{IEEEeqnarray}{rClrCl}
l^a &=& \frac{1}{\sqrt{2}} (u^a + e_r^a), & \quad
n^a &=& \frac{1}{\sqrt{2}} (u^a - e_r^a), \nonumber \\
m^a &=& \frac{1}{\sqrt{2}} (e_\theta^a + i e_\phi^a), & \quad
\bar{m}^a &=& \frac{1}{\sqrt{2}} (e_\theta^a - i e_\phi^a).
\label{eq:tetrad}
\end{IEEEeqnarray}
The normalisation in Eq. (\ref{eq:tetrad}) follows a common
convention~\cite{Baker:2001sf} used in NR and
differs from the conventions usually used in other fields.  This can
lead to constant-factor differences between expressions computed using this
convention and those appearing elsewhere in the literature.

Unfortunately, due to the degeneracy in the azimuthal coordinate,
$\phi$, this tetrad is not defined on the $z-$axis. We overcome this
problem by making the particular choice $\phi = \pi$, so that
$\phi^i = [1,0,0]$ and the tetrad is once again well defined (note, however,
that the tetrad---and hence the Weyl
scalars---remain discontinuous across the $z$-axis).

\subsection{Approximate null geodesics}

We wish to investigate the extent to which the Weyl scalars as computed
using the tetrad in Sec.~\ref{sec:tetrad} obey the peeling property
along some suitably defined curves in a BBH spacetime.  For the
peeling theorem to apply, the curves should be outgoing radial null
geodesics; this is because the Weyl scalars vary not only in advanced
time (according to the peeling theorem), but also in retarded time.  A 
failure to use exact outgoing null geodesics as the curves along which to measure the falloff
results in a mixing of these two dependencies, leading to an error in the
measured falloff.  This error may be significant or not, depending on the
extent of the deviation from exact geodesics.
It would be possible to calculate the exact null geodesics in a
numerical spacetime, but this is not typically done, and we have not
done so here as it turns out to be unnecessary for confirming the
applicability of the peeling theorem to BBH spacetimes.
Instead, we compute a series of
{\em approximate null geodesics}, defined as curves in the
$x$--$t$ coordinate plane along which local maxima of $|\Psi_4|$ propagate.
In the geometric optics approximation, these curves will be null
geodesics.

By restricting the approximate null geodesics to the $x$--$t$ plane, we
are assuming that the outgoing null geodesics which are
asymptotically radial have a negligible angular coordinate dependence
in the region in which we study them.  This is a reasonable
expectation given that in Kerr, with spin parameter equal to that of
the final post-merger black hole ($J/M^2 \approx 0.68$), the angular deviation
of an asymptotically radial outgoing null geodesic between $r=30M$ and
$r=1000M$ is only $10^{-3}$ radians.

One may expect that an even simpler approximation would suffice and make the
assumption that the null geodesics correspond approximately to those of the 
Schwarzschild spacetime in Schwarzschild coordinates. In this case the null
geodesics are given by $r_\ast = t + \text{const}$, where
$r_\ast = r + 2M \ln(r/2M-1)$ is the tortoise coordinate.
At large radius, this is not an unreasonable
expectation. However, close to the black
holes this approximation becomes increasingly poor. Unfortunately,
$\Psi_0$ and $\Psi_1$ fall off sufficiently fast that they drop below
the level of numerical error before the region $r \approx 100M$ where
the Schwarzschild approximation becomes reliable, so this approximation is unsuitable
for measuring the falloff of these scalars.

\subsection{Falloff measurement}
\label{sec:falloff-measurement}

Given the Weyl scalars computed using the coordinate tetrad (\ref{eq:tetrad})
and an
approximate null geodesic $r = \lambda(t)$, described in the previous
section, we
consider the falloff of the scalars with the coordinate radius
$r$. In order to determine a representative
value for the falloff rate, we choose an interval $r \in
[r_\text{min}, r_\text{max}]$ and perform a linear least-squares fit
of
\be
\log \left | \Psi_n(t,\lambda(t)) \right | = -p \log r + \text{const}
\ee
to determine $p$.

The appropriate choice of the interval is influenced by two
factors.
Firstly, since the peeling theorem predicts only the asymptotic behaviour of the
scalars, we desire a fitting interval at as large a radius as possible
so that the sub-leading terms are negligible.  Secondly, we find that
$\Psi_0$ and $\Psi_1$ are dominated by numerical error (visible as 
spatial oscillations at the maximum grid frequency and comparable in magnitude
to the signal) at large $r$
and large $t$, and that this limits the maximum radius at which we can fit.

\subsection{Numerical implementation}
\label{setup}

Our numerical BBH solutions were obtained using the
\code{Llama} code~\cite{Pollney:2009yz} to solve Einstein's equations
in the BSSN~\cite{Nakamura:1987zz, Shibata:1995we, Baumgarte:1998te}
formalism with the moving puncture method~\cite{Baker:2005vv,Campanelli:2005dd}
using eighth-order finite differencing. The
computational infrastructure is based on the \texttt{Cactus} framework
\cite{Goodale02a,Cactuscode:web} and the \texttt{Carpet}
\cite{Schnetter:2003rb,Schnetter:2006pg,CarpetCode:web} adaptive mesh-refinement driver,
and implements a system of multiple grid patches with data exchanged
via interpolation~\cite{Pollney:2009yz}.  This multipatch technique
allowed us to use a spherical outer grid with constant angular
resolution to best match the resolution requirements of radially
outgoing waves, leading to an outer boundary at very large radius at
only modest computational cost.  This enabled us to measure the Weyl
scalars accurately at a large radius, where they are closer to their
asymptotic form.

\begin{table}
 \begin{tabular}{cccccc}
\hline
\hline
~ $m_\text{h}/M$ & $m_\text{b}/M$ & $D/M$ & $p_r/M$ & $p_t/M$\\
\hline
~ $0.5$ & $0.47653463302$ & $6$ & $-0.005867766$ & $0.138357448$\\
\hline
\hline
\end{tabular}
\caption{Initial data parameters for the equal-mass, non-spinning
  configuration studied.  $M$ is the sum of the initial irreducible
  masses of the black holes, and $m_\text{h}$ is the irreducible mass
  of each BH.  $m_\text{b}/M$, $D/M$, $p_r/M$ and $p_t/M$ are the bare
  mass, separation, radial and tangential momenta used in the
  Bowen-York initial data prescription.}
\label{tab:initialdata}
\end{table}

We studied an equal-mass binary system with an initial separation of
$6M$ and both black holes initially non-spinning.  The configuration studied
in~\cite{Pollney:2009ut} had a larger initial
separation and hence a longer gravitational wave signal.  However, the length
of the wave signal should not be important for measuring the falloff,
so we ran a shorter simulation for computational efficiency.  We computed
standard Bowen-York~\cite{Bowen:1980yu,Brandt:1997tf} initial data
with parameters given in Table~\ref{tab:initialdata} using the
\code{TwoPunctures}~\cite{Ansorg:2004ds} code.

In our simulations, calculation of the Weyl scalars was performed by
\code{WeylScal4}, a component of the open
Einstein Toolkit~\cite{einsteintoolkit}.  We have adapted \code{WeylScal4} to utilise
the \code{Llama} multipatch framework and have
performed careful testing
against analytic results for the $I$-invariant~\cite{Baker:2001sf} in Kerr.
We have also cross-checked \code{WeylScal4} against \code{Psikadelia}~\cite{psikadelia}, another code which computes
the Weyl scalars.  Since this
code was developed independently, the agreement is strong evidence that the codes are
both correct. 

\begin{table}
 \begin{tabular}{l|ccddc|cccccc}
\hline
\hline
~Run & $h_0/M$ & $N_\mathrm{ang}$ & \multicolumn{1}{c}{$R_\mathrm{in}/M$} & \multicolumn{1}{c}{$R_\mathrm{out}/M$} & $N_\mathrm{lev}$ & $r_{\text{l}}/M$ \\
\hline
~Ma   & $0.69$  & $28$ & 39.77 & 2697.6 & $6$    & $12, 6, 3, 1.5, 0.6$ \\
~Mb   & $0.60$  & $32$ & 39.6 & 2697.6 & $6$    & $12, 6, 3, 1.5, 0.6$ \\
~Ca   & $0.69$  & - & - & 200.00 & $6$    & $12, 6, 3, 1.5, 0.6$ \\
~Cb   & $0.60$  & - & - & 200.00 & $6$    & $12, 6, 3, 1.5, 0.6$ \\
\hline
\hline
\end{tabular}
\caption{Numerical grid parameters of the BBH simulations studied. Both
  multipatch (M) and Cartesian (C) simulations were run at two resolutions
  (a and b). In all cases $h_0$ is the grid spacing on the coarsest Cartesian
  grid. In cases where
  an angular grid is used this is also equal to the radial grid spacing
  in the angular patches. $N_\mathrm{ang}$ is the number of cells in the angular
  directions in the angular patches. $R_\mathrm{in}$ and
  $R_\mathrm{out}$ are the inner and outer radii of the angular
  patches. $N_\mathrm{lev}$ is the number of refinement levels
  (including the coarsest) on the Cartesian grid, and $r_{\text{l}}$
  indicates that a cubical refinement box of side $2 r_{\text{l}}$ is
  centred on the BH on level ``l'', with level 0 being the coarsest.}
\label{tab:grid}
\end{table}

We performed simulations at two different grid spacings (the detailed
grid structures used in our runs are listed in Table \ref{tab:grid})
to assess the
effect of the numerical discretisation on the solution. Owing to high
frequency numerical reflections from the inter-patch and refinement
boundaries, we found it necessary to modify the grid structure when
computing the falloff of $\Psi_0$ and $\Psi_1$ which, being the
fastest decaying and lowest in amplitude, are the most affected by
these reflections.  In this case, we eliminated the angular grid
patches entirely
and used a purely Cartesian domain with an outer boundary causally
disconnected from the region in which we could resolve these Weyl
scalars. We found that the errors coming from refinement boundaries visible in $\Psi_0$ could be
reduced significantly by 
switching the mesh refinement algorithm used
from the standard Berger-Oliger~\cite{Berger:1984zza} scheme
to the tapered grids scheme of~\cite{Lehner:2005vc}.  This has the effect of
eliminating the errors caused by time interpolation at the
mesh-refinement boundaries.  (The tapered grid scheme when used with 8th order finite differencing
typically introduces a very high computational cost on the refined levels, but in our case
the coarse level has so many grid points that this cost increase was not significant.)

\section{Results}
\label{sec:results}

We now present the results of applying the fitting methods
to the simulations described in Sec.~\ref{setup}.

\begin{figure}
\includegraphics{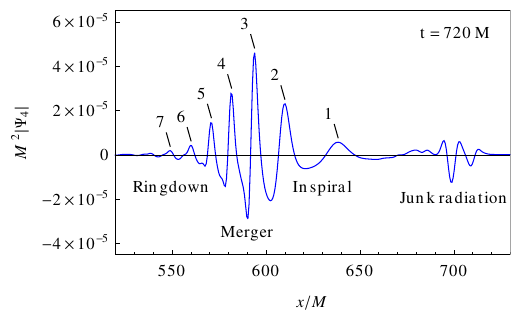}
\caption{$|\Psi_4|$ on the $x$-axis.  Note that since there is no
  multipolar decomposition, this looks different to the waveform plots
  usually shown in papers.  Specifically, the junk radiation is on the
  right and the wave propagates to the right.  The numbered labels indicate the peaks
  corresponding to the approximate null geodesics $\lambda_i$ that we compute.}
\label{fig:psi4spatial}
\end{figure}

We consider the falloff of the Weyl scalars along curves $\lambda_i$ in spacetime
which correspond to the peaks $i = 1 \ldots 7$ of $|\Psi_4|$ shown in
Fig.~\ref{fig:psi4spatial}.  This is a plot of $|\Psi_4|$ along
the $x$-axis at fixed coordinate time $t = 720M$. 
Note that for this configuration, all of the Weyl scalars
are either purely real or purely imaginary along the $x$-axis.

\begin{figure}
\includegraphics{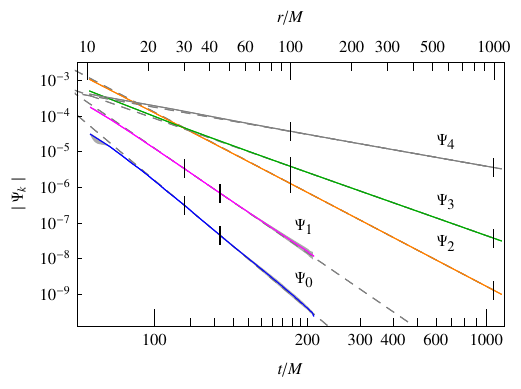}
\caption{The falloff of the Weyl scalars along the approximate null
  geodesic $\lambda_1$ corresponding to the first inspiral peak of $|\Psi_4|$ on the
  $x$-axis.  
  The falloff rates obtained from fitting in the intervals indicated by
  vertical black lines are indicated in Table \ref{tbl:results}.
}
\label{fig:falloff}
\end{figure}

Figure \ref{fig:falloff}, plotted on a log-log scale, shows the
falloff of the Weyl scalars along the curve $\lambda_1$ in the $x$--$t$ plane
corresponding to the peak labelled 1 in Fig.~\ref{fig:psi4spatial}
 as a function of the
radial coordinate $r$.  The peeling theorem predicts that the Weyl
scalars $\Psi_n$ fall off asymptotically as $r^{n-5}$
corresponding to straight lines in this figure.  The solid curves
represent the numerical data and the dashed lines represent the best
fit straight line (the line for the expected falloff is not shown, but
in each case is visually very similar to the line obtained from the
fit).  We see that to a very good approximation, each scalar exhibits a
power-law decay as predicted by the peeling theorem.

For each scalar, we choose an appropriate fitting interval (see
Sec.~\ref{sec:falloff-measurement}) and compute the falloff rate by
performing a least-squares fit in this interval.  
For \highpsis, the scalars are well resolved at all radii in our simulations
(we stopped the simulations when the signal reached $r = 1000 M$).
We therefore choose a fitting interval $r \in [100M,1000M]$ in this case.
For \lowpsis, the scalars are dominated by numerical
error beyond $r \approx 60 M$, as we observed when comparing the results at different
numerical resolutions. We
therefore choose a fitting interval $r \in [30M, 45M]$ for these.
The fitting intervals
are indicated on the plot as black vertical lines on each curve.

We have performed this analysis for each of the $7$ curves and for each
resolution.  For \lowpsis, only $\lambda_1$ and $\lambda_2$ yield meaningful
measurements of the falloff rate; there is too much finite
differencing error in the solution on $\lambda_3$--$\lambda_7$ to compute a
falloff
rate.

The shaded regions in Fig.~\ref{fig:falloff} (visible only for $\Psi_0$ 
and $\Psi_1$) represent an indication of the error due to finite differencing,
computed using the two different resolutions
and assuming eighth-order convergence of the error.  This should be
taken as indicative only, as the convergence rate obtained can vary
between $3$ and $8$ at finite resolution, corresponding to the different
sources of numerical error in the simulation.

Table \ref{tbl:results} shows the
falloff rates for each scalar along $\lambda_1$ as well as the
rate quoted in~\cite{Pollney:2009ut} and the rate expected from the peeling
theorem.  The finite differencing error in the last digit is indicated
in parentheses and again should be taken as only a coarse estimate.
The rates obtained from $\lambda_2$--$\lambda_7$ (not shown) differ by less than $2\%$ from
those obtained for $\lambda_1$ (only the curves on which the rates are sufficiently
well resolved are included).  Our measured rates are within $4\%$ of the
values expected from the peeling theorem.

\begin{table}
\begin{ruledtabular}
\begin{tabular}{c|cD{.}{.}{1.2}D{.}{.}{1.7}}
\multirow{2}{*}{$\Psi$} & \multicolumn{3}{c}{Falloff rate $p$} \\
 & Expected & \multicolumn{1}{c}{Ref.~\cite{Pollney:2009ut}} & \multicolumn{1}{c}{Measured} \\
\hline
{\rule{0pt}{2.6ex}}0 & 5 & 2.00 & 4.8(4) \\
1 & 4 & 2.48 & 3.91(7) \\
2 & 3 & 2.99 & 2.99307(6) \\
3 & 2 & 1.99 & 2.0135(6) \\
4 & 1 & 0.99 & 1.01333(7) \\
\end{tabular}

\end{ruledtabular}
\caption{Falloff rates for the Weyl scalars, including the rate expected
from the peeling theorem, the rate obtained in~\cite{Pollney:2009ut}, and the rate measured
from our simulations for the approximate null geodesic $\lambda_1$.
The error in the last digit indicated in parentheses
is a coarse estimate of the finite differencing error in the falloff rate.}
\label{tbl:results}
\end{table}

\begin{figure}
\includegraphics[width=8.71cm]{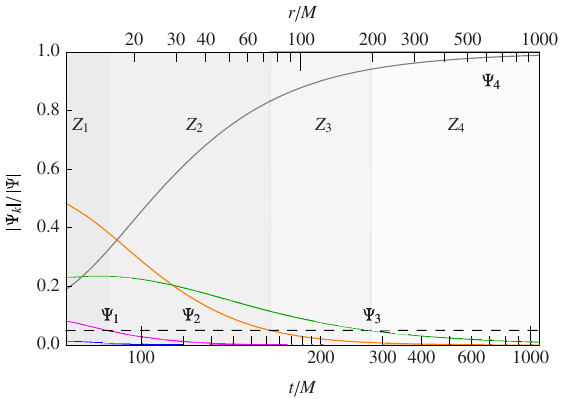}
\caption{The fractional contribution of each $\Psi_n$ (measured
  along $\lambda_1$) to the {\em total curvature}, which we define as
  $|\Psi| = \sum_{n=0}^4|\Psi_n|$.  The zones $Z_n$ are the regions in
  which $\Psi_k$, for $k \ge n$, give a contribution of more than 5\%
  to $\Psi$.  Beyond $r\gtrsim 200M$, $\Psi_4$ dominates and
  may be reliably used as a measure of gravitational radiation. Note that $Z_0$
  does not appear as $\Psi_0$ is already below $5\%$ at $r=10M$, where we can
  first start tracking the peak in $|\Psi_4|$.
}
\label{fig:regions}
\end{figure}

By studying the peeling properties of each of
the Weyl scalars we may gain insight into where $\Psi_4$ may be used as a
reasonable measure of the gravitational wave signal. This is closely related to
the identification of the regions referred to as {\em near zone}, {\em transition zones} 
and {\em radiation zone} in~\cite{Newman:1961qr}
(note that these are not the same zones referred to in~\cite{Thorne:1980ru}).
We illustrate this visually (for the curve for $\lambda_1$) in Fig.~\ref{fig:regions},
where we plot the relative contribution from each of the Weyl
scalars to the {\em total curvature}, which we define as
$|\Psi| = \sum_{n=0}^4|\Psi_n|$. Each shaded region $Z_n$ indicates the region in
which all the $\Psi_k$ with $k \ge n$ give a contribution of more than $5\%$ to the 
total curvature (notice that this also comes with a change in
the algebraic properties of the spacetime, since the principal null directions 
``peel apart'' as each $\Psi_n$ becomes important with decreasing $r$).  In
other words, as the source is approached from $r = \infty$, $Z_n$ is
the region in which $\Psi_n$ starts to make a significant contribution.
In the case of our BBH simulations, we find that $\Psi_4$ constitutes more
than $95\%$ of $|\Psi|$ in the region $r \gtrsim 200M$.
We observe that $\Psi_3$, $\Psi_2$ and $\Psi_1$ begin to contribute $> 5\%$
to the curvature at $r = 200M,75M$ and $15M$, respectively.
Note that the values of $r$ depend on the curve $\lambda_i$ along
which the falloff is measured and on the choice of cut-off percentage. For example,
the regions $Z_n$ start at lower radii for the subsequent peaks resulting in $Z_4$
beginning at $r \approx 100M$ rather than $200M$. This may be understood
from the fact that these peaks closer to the merger have stronger
gravitational wave content with the result that $\Psi_4$ is comparatively larger.

\section{Discussion}
\label{sec:discussion}

We have performed a $3$-orbit BBH simulation and measured the
falloff of the Weyl scalars, obtaining results in agreement with the peeling
theorem to within 4\%.

There are many approximations introduced in converting the precise
assumptions of the theorem into practical numerical calculations. For example,
in this work we approximated null geodesics by tracking the
location of peaks in $|\Psi_4|$ along the $x$-axis. This neglects any
angular component
in the null geodesics and also assumes that the peaks propagate along null
geodesics. Furthermore, we used a coordinate tetrad which we assume satisfies
the assumptions of the peeling theorem. The fact that we found agreement with
the predictions gives strong evidence in support of the approaches and approximations
typically used in NR simulations.

As in~\cite{Pollney:2009ut}, accurate falloff rates were easily extracted for
\highpsis, where the finite differencing error was negligible and the
deviation from the expected rate was almost certainly due to computing
the falloff at a finite radius, where sub-leading terms are nonzero.

However,
due to the presence of a large amount of numerical noise, \lowpsis~proved much more
difficult to analyse. This is not surprising given that the spacetime we are
considering---a BBH inspiral---has a strong outgoing radiation
component ($\Psi_4$ and $\Psi_3$) and Coulomb-type potential ($\Psi_2$)
and only weak incoming radiation ($\Psi_1$ and $\Psi_0$).   Nevertheless, by minimising errors coming
from inter-patch and mesh refinement boundaries, we were able to compute
\lowpsis~in a sufficiently large region to extract falloff rates within $4\%$
of the expected values. 

As discussed in Sec.~\ref{sec:results} and Fig.~\ref{fig:falloff}, the rapid
falloff of $\Psi_0$ means that it drops to a level where it is strongly
affected by numerical noise for $r \gtrsim 60M$. Furthermore, for
$r \gtrsim 100M$ it reaches a point where it is orders of magnitude below
the other $\Psi_n$ and its contribution to the curvature is negligible.
This indicates that any approximations based on the vanishing of $\Psi_0$---for 
example ``freezing-$\Psi_0$'' boundary conditions
\cite{Kidder:2004rw,Buchman:2007pj}---are robust at these large radii.

We have also shown a representative example of the radial {\em zones}
in which each of the Weyl scalars begins to contribute to the total
curvature.  In the case of our BBH simulations, we found that $\Psi_4$
constituted more than $95\%$ of $|\Psi|$ in the region $r \gtrsim
200M$.  The identification of the zones in Fig.~\ref{fig:regions}
is valid for the equal-mass, non-spinning BBH
configuration we have studied here. While we expect this to be representative
of other configurations---possibly involving spins and unequal masses---it is
likely that the exact locations of the peeling regions will be different.

The use of $\Psi_4$ to compute gravitational waveforms from NR BBH solutions 
is based on the assumption that the peeling theorem can be applied. Although it is reasonable to
expect that this is true, this work provides reassuring confirmation that this
is indeed the case. 

\begin{acknowledgments}
  It is a pleasure to thank A.~Helfer, D.~Pollney, C.~Reisswig, and E.~Schnetter
  for reading a draft of this manuscript and providing helpful
  comments and suggestions. We also thank D.~Pollney, C.~Reisswig, E.~Schnetter,
  N.~Dorband, and P.~Diener for providing the LLAMA/CTGAMMA multipatch evolution
  code used in this work, as well as all the authors of CACTUS, CARPET and the
  Einstein Toolkit for providing the open and optimized infrastructure on which
  our simulations are based. The computations were performed on the
  Datura cluster at the AEI and on the Teragrid network (allocation
  TG-MCA02N014). E.B.~acknowledges support from a Marie Curie
  International Reintegration Grant under Agreement No. PIRG05-GA-2009-249290. This work
  was supported in part by the DFG under Grant SFB/Transregio~7
  ``Gravitational-Wave Astronomy''.
\end{acknowledgments}

\bibliography{referenceseb}

\end{document}